\documentclass[11pt,a4paper]{article}
\usepackage{amsmath}

\renewcommand{\r}[1]{(\ref{#1})}

\newcommand{\xdot}{\dot{x}}
\newcommand{\dr}{\partial_r}
\newcommand{\nn}{\nonumber}
\newcommand{\go}{\gamma_{0}}

\newcommand{\alp}{\alpha}
\newcommand{\bet}{\beta}
\newcommand{\gam}{\gamma}
\newcommand{\del}{\delta}

\begin{document}

\begin{center}

{\bf\large A NEW FORM OF THE KERR SOLUTION} \\

\vspace{0.4cm}

CHRIS DORAN\footnote{e-mail: \texttt{C.Doran@mrao.cam.ac.uk}, \texttt{http://www.mrao.cam.ac.uk/$\sim$cjld1/}}

{\it Astrophysics Group, Cavendish Laboratory, Madingley Road,} \\
{\it Cambridge CB3 0HE, UK.}

\vspace{0.4cm}

\begin{abstract}
A new form of the Kerr solution is presented.  The solution involves a
time coordinate which represents the local proper time for
free-falling observers on a set of simple trajectories.  Many physical
phenomena are particularly clear when related to this time coordinate.
The chosen coordinates also ensure that the solution is well behaved
at the horizon.  The solution is well suited to the tetrad formalism
and a convenient null tetrad is presented.  The Dirac Hamiltonian in a
Kerr background is also given and, for one choice of tetrad, it takes
on a simple, Hermitian form.
\end{abstract}

\vspace{0.4cm}


\end{center}

\section{Introduction}

The Kerr solution has been of central importance in astrophysics ever
since it was realised that accretion processes would tend to spin up a
black hole to near its critical rotation rate~\cite{bar70}.  A number
of forms of the Kerr solution currently exist in the literature.  Most
of these are contained in Chandrasekhar's work~\cite{cha83}, and
useful summaries are contained in the books by Kramer~\textit{et
al.}~\cite{kra-exact} and d'Inverno~\cite{inv-rel}.  The purpose of
this paper is to present a new form of the solution which has already
proved to be useful in numerical simulations of accretion processes.
The form is a direct extension of the Schwarzschild solution when
written as
\begin{equation}
ds^2 = dt^2 - \Bigl(dr + \Bigl(\frac{2M}{r} \Bigr)^{1/2} dt \Bigr)^2 -
r^2( d\theta^2 + \sin^2\!\theta \, d\phi^2).
\label{swzmt}
\end{equation}
(Natural units have been employed.)  This is obtained from the 
Eddington-Finkelstein form
\begin{equation}
ds^2 =\Bigl(1- \frac{2M}{r}\Bigr) d{\bar{t}}^2 - \frac{4M}{r} d\bar{t}\,
dr - \Bigl(1+ \frac{2M}{r}\Bigr) dr^2 - r^2( d\theta^2
+ \sin^2\!\theta \, d\phi^2).
\end{equation}
by the coordinate transformation
\begin{equation} 
t = \bar{t} + 2 (2Mr)^{1/2}-4M \ln \Bigl( 1+
\Bigl(\frac{r}{2M}\Bigr)^{1/2} \Bigr). 
\end{equation}
In both metrics $r$ lies in the range $0 < r < \infty$, and $\theta$
and $\phi$ take their usual meaning.

The metric~\r{swzmt} has a number of nice features~\cite{DGL98-grav},
many of which extend to the Kerr case.  The solution is well-behaved
at the horizon, so can be employed safely to analyse physical
processes near the horizon, and indeed inside it~\cite{DGL98-grav}.
Another useful feature is that the time $t$ coincides with the proper
time of observers free-falling along radial trajectories starting from
rest at infinity.  This is possible because the velocity vector
\begin{equation}
\xdot^i = (1, -(2M/r)^{1/2}, 0,0)  \quad \xdot_i = (1,0,0,0)
\end{equation}
defines a radial geodesic with constant $\theta$ and $\phi$.  The
proper time along these paths coincides with $t$, and the geodesic
equation is simply
\begin{equation}
\ddot{r} = - M/r^2.
\end{equation}
Physics as seen by these observers is almost entirely Newtonian,
making this gauge a very useful one for introducing some of the more
difficult concepts of black hole physics.  The various gauge choices
leading to this form of the Schwarzschild solution also carry through
in the presence of matter and provide a simple system for the study of
the formation of spherically symmetric clusters~\cite{DDGHL-I} and
black holes~\cite{DGL98-grav}.

A further useful feature of the time coordinate in~\r{swzmt} is that
it enables the Dirac equation in a Schwarzschild background to be cast
in a simple Hamiltonian form~\cite{DGL98-grav}.  Indeed, the full
Dirac equation is obtained by adding a single term $\hat{H}_I$ to the
free-particle Hamiltonian in Minkowski spacetime.  This additional
term is
\begin{equation}
\hat{H}_I \psi = i (2M/r)^{1/2} ( \dr \psi + 3/(4r) \psi) = i
(2M/r)^{1/2} r^{-3/4} \dr (r^{3/4} \psi).
\end{equation}
A useful feature of this gauge is that the measure on surfaces of
constant $t$ is the same as that of Minkowski spacetime, so one can
employ standard techniques from quantum theory with little
modification.  One subtlety is that the Hamiltonian is not
self-adjoint due to the presence of the singularity.  This manifests
itself as a decay in the wavefunction as current density is sucked
onto the singularity~\cite{DGL98-grav}.

The time coordinate $t$ in the metric of equation~\r{swzmt} has many
of the properties of a global, Newtonian time.  This suggests that an
attempt to find an analogue for the Kerr solution might fail due to
its angular momentum.  The key to understanding how to achieve a
suitable generalisation is the realisation that it is only the local
properties of $t$ that make it so convenient for describing the
physics of the solution.  The natural extension for the Kerr solution
is therefore to look for a convenient set of reference observers which
generalises the idea of a family of observers on radial trajectories.
In Sections~\ref{S-Kerr} and~\ref{Spher} we present a new form of the
Kerr solution and show that it has many of the desired properties.  In
Section~\ref{S-tetrad} we give various tetrad forms of the solution,
and present a Hermitian form of the Dirac Hamiltonian in a Kerr
background.  Throughout we use Latin letters for spacetime indices and
Greek letters for tetrad indices, and use the signature
$\eta_{\alp\bet}=\mbox{diag ($+$ $-$ $-$ $-$)}$.  Natural units
$c=G=\hbar =1$ are employed throughout.

\section{The Kerr Solution}
\label{S-Kerr}

The new form of the Kerr solution can be written in Cartesian-type
coordinates $(t,x,y,z)$ in a manner analogous to the Kerr-Schild
form~\cite{cha83,inv-rel}.  In this coordinate system our new form of
the solution is
\begin{equation}
ds^2 = \eta_{ij} dx^i dx^j - \Bigl(\frac{2 \alp}{\rho} a_i v_j +
\alp^2 v_i v_j \Bigr) dx^i dx^j 
\label{K-cart}
\end{equation}
where $\eta_{ij}$ is the Minkowski metric,
\begin{align}
\alp &= \frac{(2Mr)^{1/2}}{\rho} \\
\rho^2 &= r^2 + \frac{a^2 z^2}{r^2},
\end{align}
and $a$ and $M$ constants.  The function $r$ is given implicitly by
\begin{equation}
r^4 - r^2(x^2+y^2+z^2-a^2) - a^2 z^2 = 0,
\end{equation}
and we restrict $r$ to $0<r<\infty$, with $r=0$ describing the disk
$z=0$, $x^2+y^2 \leq a^2$.  The maximally extended Kerr solution
(where $r$ is allowed to take negative values) will not be considered
here.

The two vectors in the metric~\r{K-cart} are
\begin{equation}
v_i = \left( 1, \frac{ay}{a^2+r^2}, \frac{-ax}{a^2+r^2},0 \right)
\label{vi}
\end{equation}
and
\begin{equation}
a_i = (r^2+a^2)^{1/2} \left( 0, \frac{rx}{a^2+r^2},
\frac{ry}{a^2+r^2}, \frac{z}{r} \right).
\label{ai}
\end{equation}
These two vectors play an important role in studying physics in a Kerr
background.  They are related to the two principal null directions
$n_{\pm}$ by
\begin{equation} 
n_{\pm} = (r^2+a^2)^{1/2} v_i \pm (\alp \rho v_i + a_i).
\label{pnd}
\end{equation} 
For computations it is useful to note that the contravariant
components of the spacelike vector in brackets are the same as those
of $-a_i$,
\begin{equation}
\alp \rho v^i + a^i = - (r^2+a^2)^{1/2} \left( 0, \frac{rx}{a^2+r^2},
\frac{ry}{a^2+r^2}, \frac{z}{r} \right).
\end{equation}
The vector $v_i$ also plays a crucial role in separating the Dirac
equation in a Kerr background, and is the timelike eigenvector of the
electromagnetic stress-energy tensor for the Kerr-Newman analogue of
our form.

\section{Spheroidal Coordinates}
\label{Spher}

The nature of the metric~\r{K-cart} is more clearly revealed if we
introduce oblate spheroidal coordinates $(r,\theta,\phi)$, where
\begin{align}
\cos\!\theta &= \frac{z}{r} \qquad 0 \leq \theta \leq \pi \\
\tan\! \phi &= \frac{y}{x}  \qquad 0 \leq \phi < 2\pi,
\end{align}
so that $\rho$ recovers its standard definition
\begin{equation}
\rho^2 = r^2 + a^2 \cos^2\theta . 
\end{equation}
The use of the symbols $r$ and $\theta$ here are standard, though one
must be aware that when $M=0$ (flat space) these reduce to
\textit{oblate spheroidal} coordinates, and not spherical polar
coordinates.  This is clear from the fact that $r$ does not equal
$\surd(x^2+y^2+z^2)$.

In terms of $(t,r,\theta,\phi)$ coordinates our new form of the Kerr
solution is
\begin{align} 
ds^2 =& dt^2 - \Bigl(\frac{\rho}{(r^2+a^2)^{1/2}} dr + \alp ( dt - a
 \sin^2\!\theta \, d\phi) \Bigr)^2 \nn \\
& - \rho^2 d\theta^2 - (r^2 +a^2) \sin^2\!\theta \, d\phi^2.
\label{NewK}
\end{align}
This neatly generalises the Schwarzschild form of equation~\r{swzmt},
replacing $\surd(2M/r)$ with $\surd(2Mr)/\rho$, and introducing a
rotational component.  The line element can be simplified further by
introducing the hyperbolic coordinate $\eta$ via $a \sinh \!\eta = r$,
though this can make some equations harder to interpret and will not
be employed here.  The metric~\r{NewK} is obtained from the advanced
Eddington-Finkelstein form of the Kerr solution,
\begin{align} 
ds^2 =& \Bigl(1 - \frac{2Mr}{\rho^2} \Bigr) dv^2 - 2\, dv \, dr +
\frac{2Mr}{\rho^2} (2 a \sin^2\!\theta) dv \, d \bar{\phi} +
2a\sin^2\!\theta \, dr \, d\bar{\phi} \nn \\
& - \rho^2 d\theta^2 - \Bigl( (r^2 +a^2) \sin^2\!\theta
+\frac{2Mr}{\rho^2} (a^2 \sin^4\!\theta) \Bigr)  d\bar{\phi}^2,
\end{align}
via the coordinate transformation
\begin{align}
dt &= dv - \frac{dr}{1+ (2Mr/(r^2+a^2))^{1/2}} \\
d \phi &= d \bar{\phi}  - \frac{a \, dr}{r^2+a^2 +
(2Mr(r^2+a^2))^{1/2}}.
\end{align}
This transformation is well-defined for all $r$, though the integrals
involved do not appear to have a simple closed form.

The velocity vector
\begin{equation}
\xdot^i = (1, -\alp(r^2+a^2)^{1/2} /\rho  , 0,0) \qquad \xdot_i = (1,
0,0,0)
\label{ff}
\end{equation}
defines an infalling geodesic with constant $\theta$ and $\phi$, and
zero velocity at infinity.  The existence of these geodesics is a key
property of the solution.  The time coordinate $t$ now has the simple
interpretation of recording the local proper time for observers in
free-fall along trajectories of constant $\theta$ and $\phi$.  As in
the spherical case, many physical phenomena are simplest to interpret
when expressed in terms of this time coordinate.  An example of this
is provided in the following section, where we show that the time
coordinate produces a Dirac Hamiltonian which is Hermitian in form.
The difference between this free-fall velocity and the velocity $v_i$
(defined by the gravitational fields) also provides a local definition
of the angular velocity contained in the gravitational field.

\section{Tetrads and the Dirac Equation}
\label{S-tetrad}

The metric~\r{NewK} lends itself very naturally to the tetrad
formalism.  From the principal null directions of equation~\r{pnd} one
can construct the following null tetrad, expressed in
$(t,r,\theta,\phi)$ coordinates, 
\begin{align}
l^i &= \frac{1}{r^2+a^2} ( r^2+a^2, \, r^2+a^2 -
\bigl(2Mr(r^2+a^2)\bigr)^{1/2}, \, 0 , \, a) \\
n^i &= \frac{1}{2\rho^2} ( r^2+a^2, \, -( r^2+a^2) -
\bigl(2Mr(r^2+a^2)\bigr)^{1/2}, \, 0 ,\, a) \\
m^i &= \frac{1}{\sqrt{2}(r+ia\cos\!\theta)} ( ia\sin\!\theta, \, 0,\, 1,\,
i \csc\!\theta).
\end{align}
In this frame the Weyl scalars $\Psi_0$,  $\Psi_1$, $\Psi_3$ and
$\Psi_4$ all vanish, and
\begin{equation}
\Psi_2 = -\frac{M}{(r-ia\cos\!\theta)^3}.
\end{equation}

A second tetrad, better suited to computations of matter geodesics, is
given by
\begin{align}
{e^0}_i &= (1,0, 0 , 0) \nn \\ 
{e^1}_i &= (\alp, \rho/(r^2+a^2)^{1/2}, 0, -\alp a \sin^2\!\theta) \nn
\\
{e^2}_i &= (0, 0, \rho, 0) \nn \\
{e^3}_i &= (0 , 0 , 0, (r^2+a^2)^{1/2} \sin\!\theta).
\end{align}
This defines a frame for all values of the coordinate $r$, so is valid
inside and outside the horizon.  Combined with the techniques
described in~\cite{DGL98-grav} this tetrad provides a very powerful
way of analysing and visualising motion in a Kerr background.

A further tetrad is provided by reverting to the original
Cartesian-type coordinates of equation~\r{K-cart} and writing
\begin{equation}
{e^\mu}_i = \del^\mu_i - \frac{\alp}{\rho} v_i a_j \eta^{j\mu},
\label{carttrad}
\end{equation}
where $v_i$ and $a_i$ are as defined at equations~\r{vi} and~\r{ai}.
The inverse is found to be
\begin{equation}
{e_\mu}^i = \del_\mu^i + \frac{\alp}{\rho} \eta^{ij} a_j \del^k_\mu
v_k.
\end{equation}
This final form of tetrad is the simplest to use when constructing the
Dirac equation in a Kerr background.  We will not go through the
details here but will just present the final form of the equation in a
Hamiltonian form.  Following the conventions of Itzykson and
Zuber~\cite{itz-quant} we denote the Dirac-Pauli matrix representation
of the Dirac algebra by $\{\gam^\mu\}$ and write $\alp^i =
\gam^0 \gam^i$, $i=1\ldots 3$.  Since ${e_\mu}^0 = \del^0_\mu$,
premultiplying the Dirac equation by $\go$ is all that is required to
bring it into Hamiltonian form.  When this is done, the Dirac equation
in a Kerr background becomes
\begin{equation}
i \partial_t \psi = -i \alp^i \partial_i \psi + m \go \psi + \hat{H}_K \psi
\end{equation}
where
\begin{align}
\hat{H}_K \psi = &  \frac{\sqrt{2M}}{\rho^2} \biggl(
(r^3 + a^2 r)^{1/4} i \dr \bigl( (r^3 + a^2 r)^{1/4} \psi \bigr) -  a
\cos\!\theta\, r^{1/4} \alp_\phi i \dr \bigl( r^{1/4}\psi \bigr) \nn \\ 
& - \frac{a\cos\!\theta}2 (r^2+a^2)^{1/2} \gam_5 \psi \biggr) 
\end{align}
and
\begin{equation}
\alp_\phi  = -\sin\!\phi\, \alp_1 +  \cos\!\phi\, \alp_2.
\end{equation}
The measure on hypersurfaces of constant $t$ is again the same as that
of Minkowski spacetime, since the covariant volume element is simply
\begin{equation}
dx\, dy \, dz = \rho^2 \sin\!\theta\, dr \, d\theta \, d\phi.
\end{equation}
As with the Schwarzschild case the interaction Hamiltonian $\hat{H}_K$
is not self-adjoint when integrated over these hypersurfaces.  This is
because the singularity causes a boundary term to be present when the
Hamiltonian is integrated.

\section{Conclusions}

The Kerr solution is of central importance in astrophysics as ever
more compelling evidence points to the existence of black holes
rotating at near their critical rate~\cite{DGL96-erice}.  Any form of
the solution which aids physical understanding of rotating black holes
is clearly beneficial.  The form of the solution presented here has a
number of features which achieve this aim.  The solution is well
suited for studying processes near the horizon, and the compact form
of the spin connection for the tetrad of equation~\r{carttrad} makes
it particularly good for numerical computation. It should also be
noted that this gauge admits a simple generalisation to a
time-dependent form which looks well-suited to the study of accretion
and the formation of rotating black holes.

A more complete exposition of the features of this gauge, including
the derivation of the Dirac Hamiltonian will be presented elsewhere.
One reason for not highlighting more of the advantages here is that
many of the theoretical manipulations which exploit these properties
have been performed utilising Hestenes' spacetime
algebra~\cite{DGL98-grav,hes-sta}.  This language fully exposes much
of the intricate algebraic structure of the Kerr solution and brings
with it a number of insights.  These are hard to describe without
employing spacetime algebra and so will be presented unadulterated in
a separate paper.

The fact that the time coordinate measured by a family of free-falling
observers brings the Dirac equation into Hamiltonian form is
suggestive of a deeper principle.  This form of the equations also
permits many techniques from quantum field theory to be carried over
to a gravitational background with little modification.  The lack of
self-adjointness due to the source itself is also natural in this
framework, as the singularity is a natural sink for the current.  In
the non-rotating case the physical processes resulting from the
presence of this sink are quite simple to analyse~\cite{DGL98-grav}.
The Kerr case is considerably more complicated, due both to the nature
of the fields inside the inner horizon, and to the structure of the
singularity.  One interesting point to note is that the sink region is
described by $r=0$, and so represents a disk, rather than just a ring
of matter.  This in part supports the results of earlier calculations
described in~\cite{DGL96-erice}, though much work remains on this
issue.

\section*{Acknowledgements}

CD is supported by an EPSRC Fellowship.  The author is grateful to
Anthony Lasenby and Anthony Challinor for helpful discussions.


\begin{thebibliography}{1}

\bibitem{bar70}
J.M. Bardeen.
\newblock Kerr metric black holes.
\newblock {\em Nature}, 226:64, 1970.

\bibitem{cha83}
S.~Chandrasekhar.
\newblock {\em The Mathematical Theory of Black Holes}.
\newblock Oxford University Press, 1983.

\bibitem{kra-exact}
D.~Kramer, H.~Stephani, M.~Mac{C}allum, and E.~Herlt.
\newblock {\em Exact Solutions of Einstein's Field Equations}.
\newblock Cambridge University Press, 1980.

\bibitem{inv-rel}
R.~d'Inverno.
\newblock {\em Introducing {E}instein's Relativity}.
\newblock Oxford University Press, 1992.

\bibitem{DGL98-grav}
A.N. Lasenby, C.J.L. Doran, and S.F. Gull.
\newblock Gravity, gauge theories and geometric algebra.
\newblock {\em Phil. Trans. R. Soc. Lond. A}, 356:487--582, 1998.

\bibitem{DDGHL-I}
A.N. Lasenby, C.J.L. Doran, M.P. Hobson, Y.~Dabrowski, and S.F. Gull.
\newblock Microwave background anisotropies and nonlinear structures {I}.
  {I}mproved theoretical models.
\newblock {\em Mon. Not. R. Astron. Soc.}, 302:748, 1999.

\bibitem{itz-quant}
C.~Itzykson and J-B. Zuber.
\newblock {\em Quantum Field Theory}.
\newblock McGraw-Hill, New York, 1980.

\bibitem{DGL96-erice}
A.N. Lasenby, C.J.L. Doran, Y.~Dabrowski, and A.D. Challinor.
\newblock Rotating astrophysical systems and a gauge theory approach to
  gravity.
\newblock In N.~S{\'{a}}nchez and A.~Zichichi, editors, {\em Current Topics in
  Astrofundamental Physics, Erice 1996}, page 380. World Scientific, Singapore,
  1997.

\bibitem{hes-sta}
D.~Hestenes.
\newblock {\em Space-Time Algebra}.
\newblock {Gordon and Breach, New York}, 1966.

\end{thebibliography}
\end{document}